# Group Evolution Discovery in Social Networks

Piotr Bródka[1,2], Stanisław Saganowski[1], Przemysław Kazienko[1,3]
[1] Institute of Informatics, Wrocław University of Technology, Wyb.Wyspiańskiego 27, 50-370 Wrocław, Poland
[2] TELNET – LOKUM S.K.A., ul. Obrońców Poczty Gdańskiej 13A 52-204 Wrocław, Poland
[3] Research Engineering Center Sp. z o.o., ul. Strzegomska 46B, 53-611 Wrocław, Poland
piotr.brodka@pwr.wroc.pl, stanislaw.saganowski@student.pwr.wroc.pl, kazienko@pwr.wroc.pl

*Abstract*—Group extraction and their evolution are among the topics which arouse the greatest interest in the domain of social network analysis. However, while the grouping methods in social networks are developed very dynamically, the methods of group evolution discovery and analysis are still '*uncharted territory*' on the social network analysis map. Therefore the new method for the group evolution discovery called *GED* is proposed in this paper. Additionally, the results of the first experiments on the email based social network together with comparison with two other methods of group evolution discovery are presented.

## I. INTRODUCTION AND RELATED WORK

Social network analysis (SNA) is gaining on importance every day, mostly because of growing number of different social networking systems and growth of the Internet. Matter of the social networking systems may be various, starting with physical system (transportation and energy networks), through virtual systems (Internet, telecommunication, WWW), social networks, biological networks, ending on food webs and ecosystems[1]. Network analysed in this paper is a social network, which in simplest form can be described as set of actors (network nodes) connected by relationships (network edges). Many researchers proposed their own concept of social network [3], [4]. [5]. [6]. Social networks, as an interdisciplinary domain, might have different form: corporate partnership networks (law partnership) [7], scientist collaboration networks [8], movie actor networks, friendship network of students [9], company director networks [10], sexual contact networks [11], labour market [12], public health [13], psychology [14], etc.

The next, section describes the related work. The third section paragraphs *A* and *B* present the few basic concepts like temporal social network, group or social position measure which help to understand the new method. Next, in paragraph *C* of the third section the group evolution and its steps are described, and in paragraph *D* the new method of group evolution extraction based on members position in social network called GED is presented. In section IV the results of experimental studies are presented followed by conclusions in section V.

## II. RELATED WORK

The easiest to investigate, social networks, are online social networks [15], [16], web-based social networks [17], computer-supported social networks [18] or virtual social networks. The reason for this is simple and continuous way to obtain data from which we can extract those social networks. Depending on the type of social network, data can be found in various places, e.g.: bibliographic data [19], blogs [20], photos sharing systems like Flickr [21], e-mail systems [22], [23], telecommunication data [24], [25], social services like Twitter [26] or Facebook [27], [28], video sharing systems like YouTube [29], Wikipedia [30] and more. Obtaining data from mentioned "data sources" allows to explore more than single social network in specific snapshot of time. Using proper techniques it is possible to evaluate changes occurring in social network over time. Especially interesting is following changes of social groups (communities) extracted from social networks.

In recent years many methods for tracking changes in social groups have been proposed. Sun et al. have introduced GraphScope [33], Chakrabarti et al. have presented original approach in [34], Lin et al. have provided framework called FacetNet [35], Kim and Han in [36] have introduced the concept of *nono-communities*, Hopcroft et al. in [37] have also investigated group evolution, however no method which can be implemented have been provided. Two methods evaluated in this article are described with more details below.

Asur et al. have proposed in [38] simple approach for investigating group evolution over time. First, groups are extracted in each timeframe, then comparing size and overlapping of every possible pair of groups in consecutive time steps events involving those groups are assigned. When none of the nodes in group from time step $i$ occur in following timeframe $i+1$, Asur et al. have described this situation as dissolve of the group. In opposite to dissolve, if none of the nodes in group from timeframe $i$ was present in previous timeframe $i–1$, group is marked as new born. Group continue its existence when identical occurrence of the group in consecutive timeframes is found. Situation when two groups from time step $i–1$ joined together overlap with more than selected percentage of the single group in timeframe $i$, is called merge. Opposite case, when two groups from timeframe $i$ joined together overlap with more than selected percentage of the single group in timeframe $i+1$, is marked as split. Asur et al. did not specify which method has been used for group extraction, nor if method works for overlapping groups.

The work was supported by: Fellowship co-Financed by European Union within European Social Fund, The Polish Ministry of Science and Higher Education, the development project, 2009-2011 and The Polish Ministry of Science and Higher Education the research project, 2010-2013

Palla et al. in [39], [40] have used clique percolation method (CPM) [41], [42], which allows group to overlap. Thanks to this feature analysing changes in groups over time is very simple. Networks at two consecutive timeframes $i$ and $i+1$ are merged into single graph $Q(i, i+1)$ and groups are extracted using CPM method. Next, the communities from timeframes $i$ and $i+1$, which are the part of the same group from joined graph $Q(i, i + 1)$, are considered to be matching. It may happen that more than two communities are contained in the same group. Then, matching is performed based on the value of their relative overlap sorted in descending order. Possible events between groups are: growth, contraction, merging, splitting, birth and death. Using CPM method, Palla et al. allowed to investigate evolution in overlapping groups, which can be extracted from directed as well as weighted network.

### III. GROUP EVOLUTION DISCOVERY

Before the method can be presented, it is necessary to describe a few concepts related to social networks

#### A. Temporal Social Network and Groups

Temporal social network *TSN* is a list of following timeframes (time windows) *T*. Each timeframe is in fact social network $SN(V,E)$ where $V$ – is a set of vertices and $E$ is a set of directed edges $<x,y>: x,y \in V$

$$\begin{aligned}
TSN &=< T_1, T_2, \ldots, T_m >, \quad m \in N \\
T_i &= SN_i(V_i, E_i), \quad i = 1, 2, \ldots, m \\
E_i &=< x, y >: x, y \in V_i, \quad i = 1, 2, \ldots, m
\end{aligned} \quad (1)$$

There is no universally acceptable definition of the groups in social networks [43], [41]. There are several of them, which are used depending on the authors' needs. In addition, some of them cannot be even called definitions but only criteria for the group existence.

A group, often also called a community, in the biological terminology is a group of cooperating organisms, sharing a common environment. In sociology, in turn, it is traditionally defined as a group of people living and cooperating in a single location. However, due to the fast growing and spreading Internet, the concept of community has lost its geographical limitations. Overall, a general idea of the social community is a group in a given population, whose members more frequently collaborate with each other rather than with other members of this population (the entire social network). The concept of the group (social community) can be easily transposed to the graph theory, in which the social network is a graph and a group as a set of vertices with high density of edges inside the group, and lower edge density between nodes belonging to two separate groups. However, the problem arises in the quantitative definition of community. Most definitions are build based on the idea presented above. Nevertheless, as mentioned earlier, there are many alternative approaches and none of them has been commonly accepted [44], [45], [46]. Additionally, groups can also be algorithmically determined, as the output of the specific clustering algorithm, i.e. without a precise a priori definition [47]. In this paper, we will use such definition, i.e. a group $G$ extracted from the social network $SN(V,E)$ is a subset of vertices from $V$ ($G \subseteq V$), extracted using any community extraction method (clustering algorithm).

#### B. Group Evolution

Group evolution is a sequence of events (changes) succeeding each other in the successive time windows (timeframes) within the social network. Possible events in social group evolution are, see Figure 1:

1. *Continuing* (stagnation), when groups in the consecutive time windows are identical or when groups differ only by few nodes and their size remains the same.

2. *Shrinking*, when nodes has left the group, making its size smaller than in the previous time window. Like in case of growing, a group can shrink slightly as well as greatly.

3. *Growing* (opposite to shrinking), when new nodes has joined to the group, making its size bigger than in the previous time window. A group can grow slightly as well as significantly, doubling or even tripling its size.

4. *Splitting* occurs, when a group splits into two or more groups in the next time window. Like in merging, we can distinguish two types of splitting: equal and unequal, which might be similar to shrinking.

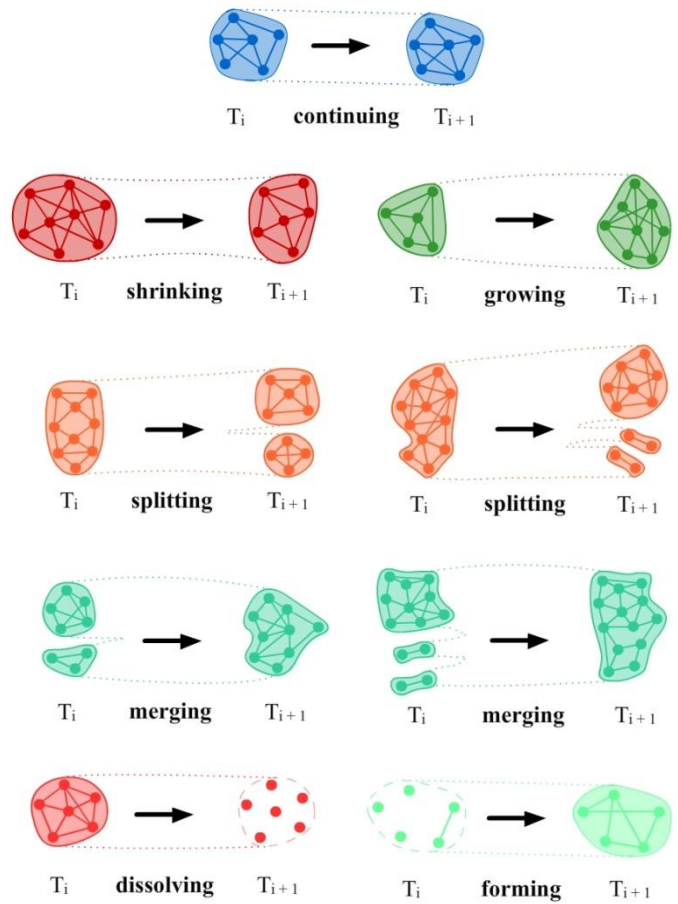

Figure 1. The events in group evolution

5. *Merging*, (reverse to splitting) when a group consist of two or more groups from the previous time window. Merge might be (1) *equal*, which means the contribution of the groups in merged group is almost the same, or (2) unequal, when one of the groups has much greater contribution into the merged group. In second case merging might be similar to growing.

6. *Dissolving*, when a group ends its life and does not occur in the next time window.

7. *Forming* (opposed to dissolving) of new group, which has not exist in the previous time window. In some cases, a group can be inactive over several timeframes, such case is treated as dissolving of the first group and forming again of the second one.

C. *Social Position*

The *GED* (see Section *III D*) method, to discover group evolution, takes into account both, the quantity and quality of the group members. To express group members quality one of the centrality measures may be used, namely, social position SP measure [31].

The social position for network $SN(V,E)$ is calculated in the iterative way, that means:

$$SP_{n+1}(x) = (1-\varepsilon) + \varepsilon \cdot \sum_{y \in V} SP_n(y) \cdot C(y \to x), \quad (2)$$

where $SP_{n+1}(x)$ and $SP_n(x)$ is the social position of member $x$ after the $n+1^{st}$ and $n^{th}$ iteration, respectively, and $SP_0(x)=1$ for each $x \in V$; $\varepsilon$ is the coefficient from the range (0;1); $C(y \to x)$ is the commitment function which expresses the strength of the relation from $y$ to $x$. For detailed information about social position measure, how to calculate and implement it see [31], [32].

D. *GED – a Method for Group Evolution Discovery in the Social Network*

To be able to evaluate the evolution of groups, a useful measure – *inclusion* of one group in another, needs to be defined. Hence, inclusion of group $G_1$ in group $G_2$ is calculated as follows:

$$I(G_1, G_2) = \frac{|G_1 \cap G_2|}{|G_1|} \cdot \frac{\sum_{x \in (G_1 \cap G_2)} SP_{G_1}(x)}{\sum_{x \in (G_1)} SP_{G_1}(x)} \quad (3)$$

Of course, instead of social position any other measure can be used e.g. centrality degree, betweenness degree, page rank etc.. However, after analysing the complexity of computation and diversity of results [32] of measures authors have decided to utilize social position measure.

As it was mentioned before the *GED* method, used to discover group evolution, takes into account both the quantity and quality of the group members. The quantity is reflected by the first part of the *inclusion* measure, i.e. what portion of their members share both groups $G_1$ and $G_2$, whereas the quality is expressed by the second part of the *inclusion* measure, namely what contribution of important members share groups $G_1$ and $G_2$. It provides a balance between the groups, which contain many of the less important members and groups with only few but key members.

It is assumed that only one event may occur between two groups $(G_1, G_2)$ in the consecutive timeframes, however one group in timeframe $T_i$ may have several events with different groups in $T_{i+1}$.

---

**GED – Group Evolution Discovery Method**

1. At each timeframe $T_i$ groups are extracted from the network and social positions is counted for each user in each extracted group.

2. For each pair of groups $<G_1, G_2>$ in consecutive timeframes $T_i$ and $T_{i+1}$ inclusion of $G_1$ in $G_2$ and $G_2$ in $G_1$ is counted according to equations (3).

3. Based on inclusion and size of two groups one type of event may be assigned:

    a. *Continuing*: $I(G_1,G_2) \geq \alpha$ and $I(G_2,G_1) \geq \beta$ and $|G_1| = |G_2|$

    b. *Shrinking*: $I(G_1,G_2) \geq \alpha$ and $I(G_2,G_1) \geq \beta$ and $|G_1| > |G_2|$ OR $I(G_1,G_2) < \alpha$ and $I(G_2,G_1) \geq \beta$ and $|G_1| \geq |G_2|$ and there is only one match (matching event) between $G_2$ and all groups in the previous time window $T_i$

    c. *Growing*: $I(G_1,G_2) \geq \alpha$ and $I(G_2,G_1) \geq \beta$ and $|G_1|<|G_2|$ OR $I(G_1,G_2) \geq \alpha$ and $I(G_2,G_1) < \beta$ and $|G_1| \leq |G_2|$ and there is only one match (matching event) between $G_1$ and all groups in the next time window $T_{i+1}$

    d. *Splitting*: $I(G_1,G_2) < \alpha$ and $I(G_2,G_1) \geq \beta$ and $|G_1| \geq |G_2|$ and there is more than one match (matching events) between $G_2$ and all groups in the previous time window $T_i$

    e. *Merging*: $I(G_1,G_2) \geq \alpha$ and $I(G_2,G_1) < \beta$ and $|G_1| \leq |G_2|$ and there is more than one match (matching events) between $G_1$ and all groups in the next time window $T_{i+1}$

    f. *Dissolving*: for $G_1$ in $T_i$ and each group $G_2$ in $T_{i+1}$ $I(G_1,G_2) < 10\%$ and $I(G_2,G_1) < 10\%$

    g. *Forming*: for $G_2$ in $T_{i+1}$ and each group $G_1$ in $T_i$ $I(G_1,G_2) < 10\%$ and $I(G_2,G_1) < 10\%$

---

The scheme which facilitate understanding of the event selection for the pair of groups in the method is presented in Figure 2.

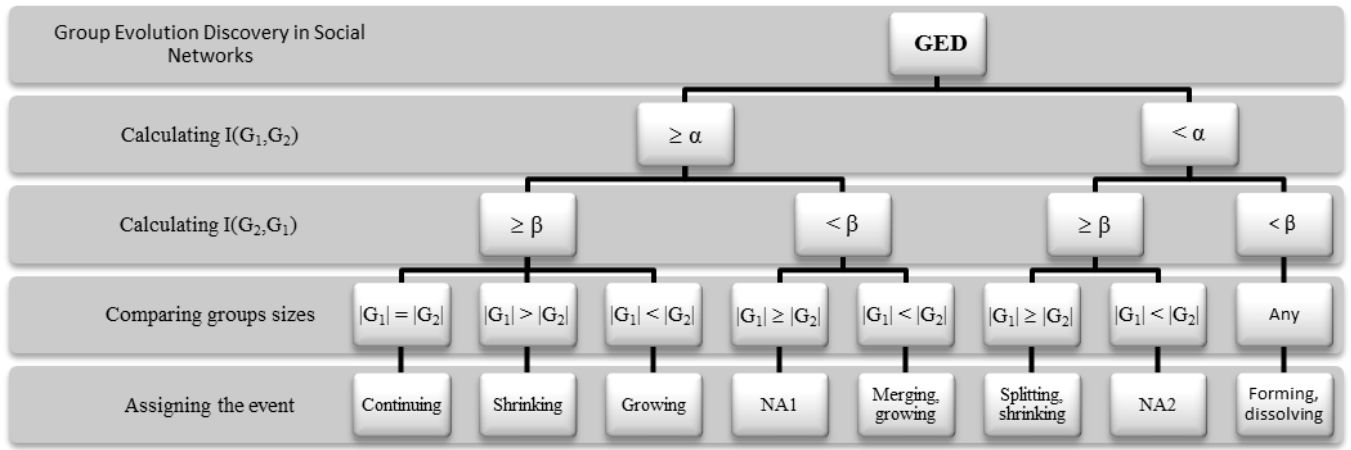

Figure 2. The decision tree for assigning the event type to the group. NA1 and NA2 are rare cases with no interpretation

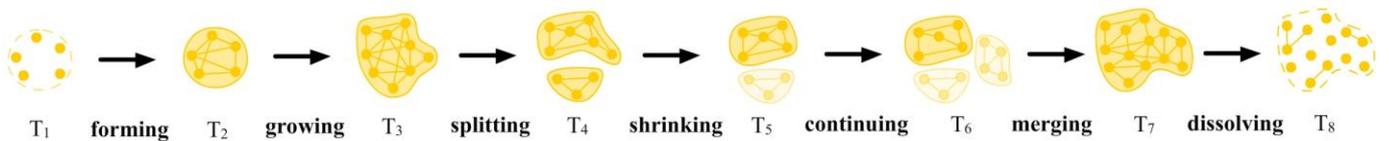

Figure 3. Changes over time for the single group.

The indicates α and β are the *GED* method parameters which can be used to adjust the method to particular social network and community detection method. After the experiments analysis (section IV) authors suggest that the values of α and β should be from range [50%;100%]

Based on the list of extracted events, which have occurred for selected group between each two successive timeframes, the group evolution is created.

In the example pictured on Figure 3 the network consists from eight time windows. The group forms in $T_2$, then by gaining new nodes grows in $T_3$, next splits into two groups in $T_4$, then by losing one node the bigger group is shrinking in $T_5$, both groups continue over $T_6$, next both groups merges with the third group in $T_7$, and finally the group dissolves in $T_8$.

## IV. EXPERIMENTS

### A. Data Set

The experiments were conducted on the data gathered from Wroclaw University of Technology email communication The whole data set was collected within period of February 2006 – October 2007 and consists of 5845 nodes and 149,344 edges.

The temporal social network consisted from fourteen 90-days timeframe was extracted from this source data. Timeframes are overlapping, the 45-days overlap, i.e., the first timeframe begins on the 1st day and ends on the 90th day, second begins on the 46th day and ends on the 135th day and so on.

For each edge in each timeframe its weight was calculated as follows:

$$w(x, y, T_i) = \frac{N(x, y, T_i)}{N(x, T_i)},$$

where: $N(x,y,T_i)$ is the number of emails sent by user $x$ to user $y$ in timeframe $T_i$ and $N(x,T_i)$ is the total number of all emails sent by user $x$ in timeframe $T_i$.

For group extraction the CPM clustering method implemented in CFinder (www.http://cfinder.org/) was utilized. CFinder has extracted from 80 to 136 groups for the timeframe (avg. 112 per timeframe). The groups was discovered for k=6 and for the directed and unweighted social network.

### B. Asur Method

First, the method introduced by Asur et al. in [38] have been implemented. As mentioned before, the CFinder method have been used for the group extraction. Afterwards all fourteen time windows have been examined, with overlapping threshold for merge and split equal 50% the authors of the method suggested 30% or 50% as a threshold. The total number of events found by Asur et al. method was 1,664, from which 90 were continuing, 72 were forming, 113 were dissolving, 703 were merging and 686 were splitting. The time needed for calculations of all time windows on home PC (2GB RAM, Intel Dual Core 1,7GHz) was more than 5.5 hours.

Such small number of continuing events is caused by very rigorous condition, which requires for groups to remain unchanged. Small amount of forming (dissolving) events came from another strong condition, which state that none of the nodes from the considered group can exist in network at previous (following) time window. A huge number of merging (splitting) events results of low overlapping threshold. In many cases to a single group from timeframe *i* has been assigned more than one type of events, e.g. group no. 1 in timeframe no.

1 was continuing in group no. 2 in timeframe no. 2 and also merging with group no. 13 from time step no. 1 into group no. 2 in timeframe no. 2. Such a case should not appear when condition for continuing events is so rigorous. However, probably the main cause of these anomalies is that Asur et al. method is not suitable for overlapping communities.

The total number of anomalies is 128 cases, 8% of all results. More than a half of these cases are groups with *split* and *merge* event into another group at the same time. The rest of the cases are even worse, because one group has *continue* and *split* or *merge* event into another group simultaneously.

### C. Palla Method

The method delivered by Palla et al. requires extraction of groups from joined graph $Q(i, i+1)$ of two consecutive timeframes $i$ and $i+1$. The first difficulty, were recognized, was extraction of groups from the large data set. Most of the joined graphs could not be grouped on home PC (2 GB RAM, Intel Dual Core 1,8GHz), and the biggest graph could not be grouped even on PC with greater computational power (8 GB RAM, Intel Core 2 Duo 2,8GHz).

Therefore, the matching algorithm by Palla et al. was tested only on the part of the data set – from timeframe no. 8 to time step no. 12. However, obtained results did not fulfil the specification provided by the authors of Palla et al. method, because only around half of the groups from timeframe $i$ and $i+1$ had its' occurrence in joined graph $Q(i, i+1)$. For this reason, the method by Palla et al. will not be taken into account at the moment while comparing results. Further attempts to compare the method will be made after consultation with the authors of the method.

### D. GED Method

A new *GED* method (see Section III *D*) was implemented using T-SQL language and was run for separately $α=(50,60,70,80,90,100)$ and $β=(50,60,70,80,90,100)$ the results are presented in Table 1.

It has to be noticed that two rare cases was identified which are not yet interpreted, i.e., NA1 and NA2 see Figure 2 and Table 1

While analysing Table 1 it can be observed that with the increase of $α$ and $β$ the number of merging events also increase while the number of shrinking, growing, splitting and continuations events decrease. So having both parameters is an advantage because the results can be adjusted to ones needs.

TABLE 1. THE RESULTS OF *GED* COMPUTATION

| α | β | exec time | \multicolumn{9}{c}{Number of different events aggregated for all 14 timeframes} | Total |
|---|---|---|---|---|---|---|---|---|---|---|---|---|
| | | | forming | dissolving | shrinking | growing | continuing | splitting | merging | NA1 | NA2 | Total |
| 50 | 50 | 05:53 | 122 | 186 | 204 | 180 | 124 | 517 | 398 | 3 | 0 | 1734 |
| 50 | 60 | 05:55 | 122 | 186 | 203 | 173 | 119 | 463 | 405 | 7 | 0 | 1678 |
| 50 | 70 | 06:32 | 122 | 186 | 197 | 157 | 112 | 398 | 421 | 19 | 0 | 1612 |
| 50 | 80 | 05:57 | 122 | 186 | 183 | 149 | 96 | 306 | 429 | 51 | 0 | 1522 |
| 50 | 90 | 06:17 | 122 | 186 | 161 | 154 | 91 | 269 | 424 | 79 | 0 | 1486 |
| 50 | 100 | 05:54 | 122 | 186 | 154 | 156 | 90 | 251 | 422 | 87 | 0 | 1468 |
| 60 | 50 | 05:56 | 122 | 186 | 190 | 177 | 124 | 531 | 359 | 0 | 0 | 1689 |
| 60 | 60 | 05:52 | 122 | 186 | 191 | 170 | 119 | 475 | 366 | 1 | 0 | 1630 |
| 60 | 70 | 05:52 | 122 | 186 | 187 | 152 | 112 | 408 | 384 | 8 | 0 | 1559 |
| 60 | 80 | 05:51 | 122 | 186 | 178 | 144 | 96 | 311 | 392 | 33 | 0 | 1462 |
| 60 | 90 | 05:51 | 122 | 186 | 159 | 148 | 91 | 271 | 388 | 54 | 0 | 1419 |
| 60 | 100 | 05:52 | 122 | 186 | 152 | 149 | 90 | 253 | 387 | 60 | 0 | 1399 |
| 70 | 50 | 06:58 | 122 | 186 | 179 | 173 | 123 | 543 | 281 | 0 | 6 | 1613 |
| 70 | 60 | 05:50 | 122 | 186 | 180 | 169 | 119 | 486 | 285 | 0 | 2 | 1549 |
| 70 | 70 | 05:49 | 122 | 186 | 177 | 156 | 112 | 418 | 298 | 1 | 0 | 1470 |
| 70 | 80 | 05:49 | 122 | 186 | 173 | 149 | 96 | 316 | 305 | 17 | 0 | 1364 |
| 70 | 90 | 05:55 | 122 | 186 | 157 | 150 | 91 | 273 | 304 | 32 | 0 | 1315 |
| 70 | 100 | 06:33 | 122 | 186 | 150 | 152 | 90 | 255 | 302 | 36 | 0 | 1293 |
| 80 | 50 | 05:56 | 122 | 186 | 172 | 148 | 120 | 553 | 230 | 0 | 24 | 1555 |
| 80 | 60 | 06:27 | 122 | 186 | 173 | 144 | 117 | 495 | 234 | 0 | 11 | 1482 |
| 80 | 70 | 06:23 | 122 | 186 | 170 | 134 | 111 | 426 | 244 | 0 | 3 | 1396 |
| 80 | 80 | 07:17 | 122 | 186 | 165 | 127 | 96 | 324 | 251 | 1 | 0 | 1272 |
| 80 | 90 | 06:34 | 122 | 186 | 154 | 128 | 91 | 276 | 250 | 8 | 0 | 1215 |
| 80 | 100 | 06:40 | 122 | 186 | 148 | 129 | 90 | 257 | 249 | 10 | 0 | 1191 |
| 90 | 50 | 06:35 | 122 | 186 | 172 | 134 | 120 | 553 | 188 | 0 | 46 | 1521 |
| 90 | 60 | 06:28 | 122 | 186 | 174 | 131 | 117 | 494 | 191 | 0 | 28 | 1443 |
| 90 | 70 | 07:29 | 122 | 186 | 171 | 125 | 111 | 425 | 197 | 0 | 9 | 1346 |
| 90 | 80 | 06:41 | 122 | 186 | 165 | 120 | 96 | 324 | 202 | 0 | 2 | 1217 |
| 90 | 90 | 06:01 | 122 | 186 | 154 | 123 | 91 | 276 | 199 | 0 | 0 | 1151 |
| 90 | 100 | 06:28 | 122 | 186 | 148 | 123 | 90 | 257 | 199 | 1 | 0 | 1126 |
| 100 | 50 | 07:24 | 122 | 186 | 176 | 116 | 120 | 549 | 172 | 0 | 64 | 1505 |
| 100 | 60 | 07:43 | 122 | 186 | 177 | 114 | 117 | 491 | 174 | 0 | 44 | 1425 |
| 100 | 70 | 06:06 | 122 | 186 | 173 | 111 | 111 | 423 | 177 | 0 | 19 | 1322 |
| 100 | 80 | 06:34 | 122 | 186 | 166 | 111 | 96 | 323 | 177 | 0 | 7 | 1188 |
| 100 | 90 | 06:15 | 122 | 186 | 154 | 115 | 91 | 276 | 173 | 0 | 2 | 1119 |
| 100 | 100 | 06:13 | 122 | 186 | 148 | 115 | 90 | 257 | 173 | 0 | 0 | 1091 |

So the choice of proper group extraction algorithm should be considered. If one needs overlapping groups for a small network then CPM can be used, but if one needs to extract groups very fast and for a big network than the method proposed by Blondel [49] can be utilized. That is a big advantage because most method can be used only for either overlapping or non-overlapping groups.

The next advantage of the *GED* method is the fact that any user's measure which describes user's importance can by utilized. If someone need results very fast or does not possess enough computational power then the simple degree centrality can be used, but if there is both time and computational power then some more complex measures like betweenness or page rank may be utilized. Moreover the *GED* method will work even without any measure (just 1 for each node), but then the results might be similar to method proposed by Asur et al.

All the advantages described above make the *GED* method extremely flexible. It can analyse both, small and very large social networks, can return results very fast as well as analyse the network deeper.

As regards the method by Asur et al., the *GED* method is much faster. As already mentioned, the computation time for Asur was more than 5.5 hours while for *GED* it took less than 4 hours to calculate events for $\alpha$=(50,60,70,80,90,100) and $\beta$=(50,60,70,80,90,100).

Another great advantage over Asur method is that the *GED* method can be successfully used for overlapping communities, while the method by Asur et al. generates abnormal results, see Section IV*C*.

Last but not least, another difference is that the *GED* method allows to change inclusion thresholds, which effect on the methods strictness and allow to adjust the results to user's needs.

Comparing to method by Palla et al. the *GED* method is performing incomparable faster. As mentioned before the grouping with CFinder was unsuccessful for two timeframes. For timeframes which have been grouped correctly, execution time ranged from 5 minutes to 16 hours using better PC. Since for each two timeframe it was necessary to extract groups three times (to extract groups from timeframes $T_i$, $T_{i+1}$ and their union) it took a lot of time to define events between groups.

Additionally Palla's et al. method requires usage of *CPM* method, while the *GED* method may be applied for all existing group extraction algorithms.

## V. CONCLUSIONS

The increasing number of systems in which people communicate with each other continue to rise. That creates an insatiable need to analyse them. One part of such analysis is groups extraction and analysis of their evolution over time in order to understand the mechanisms governing the development and variability of social groups.

The *GED* method, proposed in the paper, uses not only the size and comparison of groups members, but also takes into account their position and importance in the group to determine what happened with the group in successive timeframes.

The *GED* method was designed to be as much flexible as possible and fitted to both, overlapping and non-overlapping groups but also to has low and adjustable computational complexity.

The results of first experiments and comparison with the existing methods presented in section IV, leads to the conclusion that desired features were achieved, and the new *GED* method may become one of the best method for group evolution discovery.

## VI. FUTURE WORK

The method presented in this paper is still in the development phase. The next step will be to add possibility to detect migration, i.e., if the group has split into two or more groups we would like to know to which group the core of given group (the most important members - leaders) have migrated and which part of the group followed the leaders.

Afterward, the method will be tested on as many group extraction method as possible and compared to other group evolution discovery methods.

Additionally the adjustment of the method to work with multi-layered social networks will be performed